**Mechanical Isolation of Highly Stable Antimonene under Ambient Conditions**

By *Pablo Ares*, *Fernando Aguilar-Galindo*, *David Rodríguez-San-Miguel*, *Diego A. Aldave*, *Sergio Díaz-Tendero*, *Manuel Alcamí*, *Fernando Martín*, *Julio Gómez-Herrero\** and *Félix Zamora\**

P. Ares, D. A. Aldave, Prof. J. Gómez-Herrero
Departamento de Física de la Materia Condensada
Universidad Autónoma de Madrid, Madrid E-28049 (Spain)
E-mail: julio.gomez@uam.es

F. Aguilar-Galindo, Dr. S. Díaz-Tendero, Prof. M. Alcamí, Prof. F. Martín
Departamento de Química
Universidad Autónoma de Madrid, Madrid E-28049 (Spain)

D. Rodríguez-San-Miguel, Dr. F. Zamora
Departamento de Química Inorgánica
Universidad Autónoma de Madrid, Madrid E-28049 (Spain)
E-mail: felix.zamora@uam.es

Dr. S. Díaz-Tendero, Prof. F. Martín, Prof. J. Gómez-Herrero, Dr. F. Zamora
Condensed Matter Physics Center (IFIMAC)
Universidad Autónoma de Madrid, Madrid E-28049 (Spain)

Prof. M. Alcamí, Prof. F. Martín, Dr. F. Zamora
Instituto Madrileño de Estudios Avanzados en Nanociencia (IMDEA-Nanociencia)
Cantoblanco, Madrid E-28049 (Spain)

The extraordinary success of graphene and its tremendous potential applications[1] paved the way for the rising of a completely new family of two dimensional materials.[2] Graphene is a semimetal with zero-gap, which limits its use in the electronics technology. Transition metal dichalcogenides present a band gap in the range of 1.5 - 2.5 eV[3] (depending on the thickness, strain level and chemical composition), which makes them inappropriate for some optoelectronics applications where band gaps in the 0.1 - 1 eV range are commonly preferred.[4] Black phosphorous (BP),[5] a layered allotrope of phosphorous, presents an energy gap in this range and hence it is now intensely studied to better understand its electronics properties in the few-layer conformation. However, it shows a relatively large reactivity. Exfoliated flakes of BP are highly hygroscopic and tend to uptake moisture from air. The long term contact with water condensed on the surface seems to degrade BP, as it can be seen from measurements of flake topography over several hours,[6] by measuring the electrical performance of transistors or the sheet resistance as a function of time. Phosphorus belongs to the nitrogen group (group 15 in the periodic table of elements). In this same group, we also find antimony, a silvery lustrous, non-hygroscopic, gray metal with a layered structure similar to that of BP. Theoretical calculations[7] point out towards an electronic structure with a band gap suitable for optoelectronics applications. In this work, we report both micromechanical exfoliation of antimony down to the single-layer regime and experimental evidence of its stability.[8] Our experiments demonstrate that single/few-layer antimony flakes are highly stable in ambient conditions showing mechanical stability upon origami nanomanipulation and no degradation over month periods. Density functional theory (DFT) simulations mimicking ambient conditions confirm the geometrical experimental findings and predict a band gap of 1.2-1.3 eV within the range of optoelectronics applications. The procedure to prepare single and few-layer flakes started by mechanical exfoliation of a macroscopic freshly cleaved crystal of antimony (**Fig. 1**). As usual, submillimeter flakes were easily obtained by repetitive pealing using adhesive tape (Fig. 1a). Attempts to transfer directly these flakes to a substrate (300 nm oxide silicon grown thermally on a Si (111) crystal) by pressing the tape against the substrate, as is widely done with other 2D materials, resulted in a very low transfer yield of thick flakes with many layers. We then adopted a more sophisticated strategy consisting of an initial transfer from the adhesive tape to a thin layer of viscoelastic polymer (Gelfilm® from Gelpak®) which was adhered to a glass slide to facilitate its handling.[9] The softness of the viscoelastic polymer results in a higher yield of flakes on the polymer surface. Then, by pressing the polymer against the silicon oxide substrate, we were able to obtain thin antimony flakes with large areas in a more controlled way as seen by optical microscopy (Fig. 1b, Fig. S1). Atomic force microscopy (AFM) revealed smaller crystals with thickness in the nanometer range (Fig. 1c, Fig. S2). Fig. 1d is an AFM topography showing a ~ 0.2 µm² flake with different terraces. According to the height histogram and profile (Fig. S2) the minimum layer thickness is ~ 1.8 nm compatible with a bilayer of antimony. Importantly, the antimony flakes did not show any measurable Raman signal for thickness below ~100 nm (Fig. S3). This behavior is observed in other layered materials such as mica[10] and has no implications upon stability.

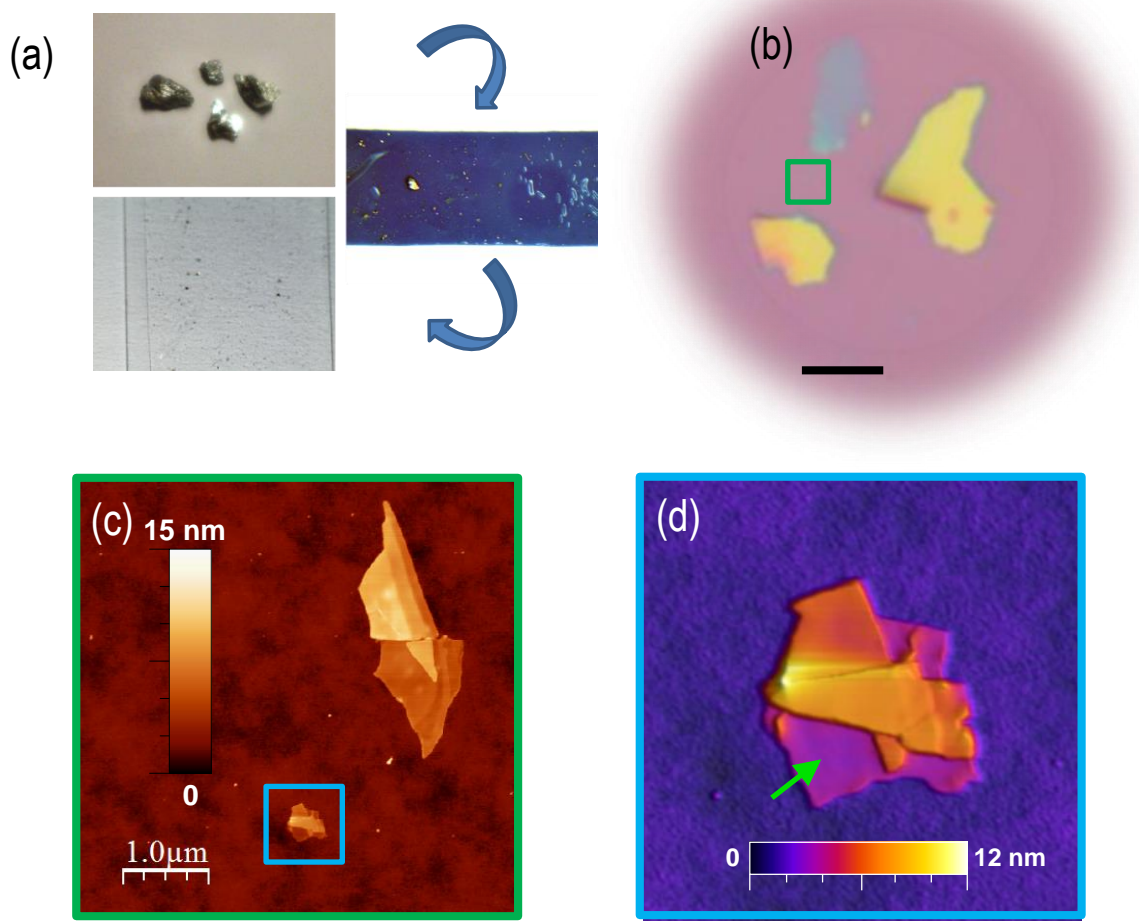

**Figure 1.** Antimonene flakes on SiO2 substrates. (a) Top left, millimeter size crystals of antimony. Middle right, adhesive tape with submillimeter crystals of antimony. Bottom left, polymer on glass slide with micrometer antimony flakes. (b) Optical microscopy image where up to 3 large flakes of antimony can be seen. Black scale bar 10 µm. Different colors reflect different thickness. (c) AFM topographic image showing 2 flakes of antimonene located inside the marked region in (b). (d) AFM topography of the ~ 0.2 µm2 antimonene flake inside the blue square in (c) showing terraces of different heights. The terrace with minimum thickness (compatible with a bilayer, see Fig. S2) is marked with a green arrow.

**Fig. 2** depicts a detailed study of the structure of thin antimonene flakes. Fig. 2a displays relevant views and parameters of the antimony atomic lattices. Fig. 2b shows a high resolution Transmission Electron Microscopy (TEM) image of a few layer antimonene flake. The image reveals very thin well-resolved terraces. The atomic structure from the different layers shows a clear hexagonal periodicity (see inset) that corresponds to that expected for few layer antimonene (β-Sb phase).

Fig 2c displays a high-resolution AFM topographic image taken in the lowest terrace of the isolated flake shown in Fig. 1d nearby the spot marked by the green arrow. The image exhibits an atomic periodicity again compatible with that expected for antimony (Fig. 2d). This image was acquired after exposing the flake to atmospheric conditions during more than two months.

Additional TEM measurements confirm these results (Fig. S4). Electron diffraction data are in good agreement with the Fourier transforms of high-resolution AFM topographic images (Fig. S4). The composition of the isolated flakes was corroborated by X-Ray Energy Dispersive Spectroscopy (XEDS) microanalysis (Fig. S5).

DFT simulations have been performed considering one and two monolayers (ML) of antimony. In order to mimic the experimental conditions, simulations have been carried out at room temperature and also by including solvent effects (water and oxygen environments, see Supporting Information). As a reference, they have also been performed at 0 K in vacuum. Fig. 2d shows the geometry obtained in the case of one and two monolayers of antimony. The simulations predict a hexagonal order (top view) with different heights for the atoms (lateral views), in good agreement with TEM and AFM measurements (Figs. 2b-c).

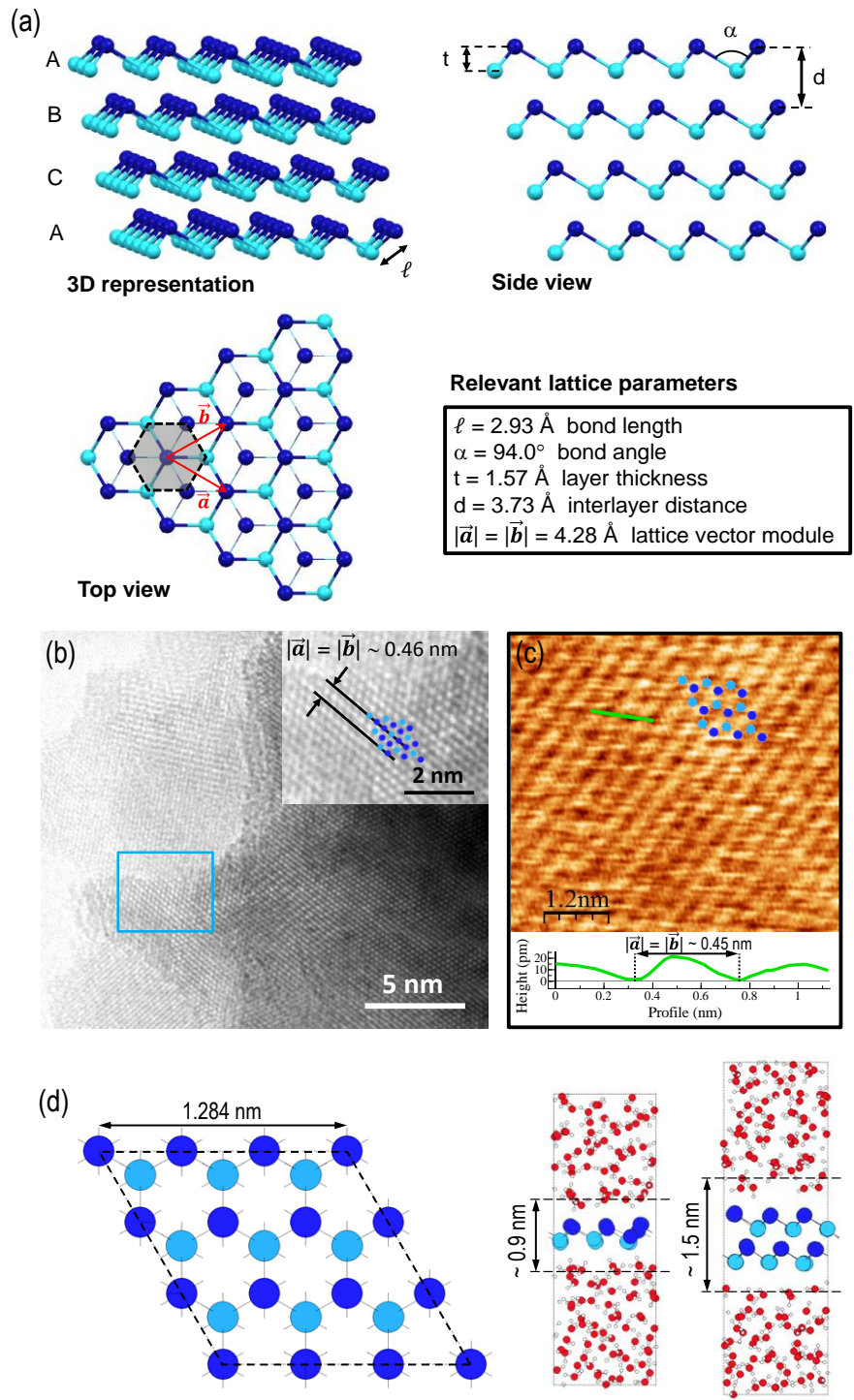

**Figure 2.** (a) Relevant views and parameters of antimony atomic lattice. (b) High resolution TEM image of a few layer antimonene flake. The inset is a digital magnification of the area inside the blue rectangle. Bilayer antimonene structure is superimposed showing a good agreement with the hexagonal lattice. (c) AFM topography acquired on the bilayer terrace marked with a green arrow in Fig. 1d showing atomic periodicity. The superimposed single layer antimonene atomic lattice is compatible with the observed periodicity (dark blue circles correspond to top atom positions and light blue circles to bottom ones). Profile on the bottom was taken along the green line in the image. The measured distance for the lattice vector module is in good agreement within experimental precision with the crystallographic structure. (d) Left: top view atomic lattice of antimonene. Right: side view of mono and bilayer antimonene lattices including water molecules as used for DFT calculations.

In simulations performed at room temperature, we have already observed a slight deformation of the crystal structure (~ 5 % in Sb-Sb distances) in comparison with the case in vacuum. The presence of water or oxygen barely affects these geometries.[11] The distance predicted between the centers of the second closest upper and lower water molecules is ~ 0.9 and ~ 1.5 nm for the monolayer and bilayer cases, respectively. Concerning the electronic structure, we have computed the density of states (DOS) for 1 and 2 ML in vacuum and at room temperature in a water environment (see Supporting Information). Our prediction for the gap in vacuum for 1ML is 1.6 eV, which is in good agreement with previous calculations.[7b] We also observe that the band gap closes when going from 1 to 2 ML,[7a, 7b] where the antimonene shows metallic character. More importantly, our calculations mimicking ambient conditions lead to a gap reduction from 1.6 at 0 K to 1.3 eV at room temperature in the 1ML case, and down to 1.2 eV when the system is placed in a water environment. The presence of oxygen does not alter these figures.

**Fig. 3** depicts a detailed characterization of a single antimonene layer. Fig. 3a shows a few layer antimonene flake with a well-defined monolayer terrace located at its bottom. As for graphene, rippling is caused by conformation of antimonene to the underlying $SiO_2$, and is not intrinsic.[12] Rippling makes very difficult to obtain well-resolved AFM images of atomic periodicity. The measured height of this terrace is ~ 0.9 nm (Fig. 3a-b) compatible with the presence of water layers as shown in Fig. 2d. It is widely assumed that under ambient conditions there exists an ever–present layer of adsorbed water (with a thickness of ~ 0.6 nm) which remains captured between the flakes and $SiO_2$.[13]

To determine whether this terrace is a monolayer nanomanipulation with AFM was performed, folding the layer into an origami structure (Fig. 3c). According to Geim and Novoselov[13a] the identification of single graphene sheets can be unambiguously carried out by measuring the step height of single folds. In our case, the lowest step height is ~ 0.4 nm (inset Fig. 3c and profile Fig. 3d) that corresponds to a single layer of antimonene. Furthermore, this origami nanomanipulation was performed several days after flake deposition on the substrate. The folding of the sheet itself and the angles observed in the origami structure (mainly multiples of 60 °, expected from a hexagonal atomic structure) show the mechanical stability of single antimonene sheets.

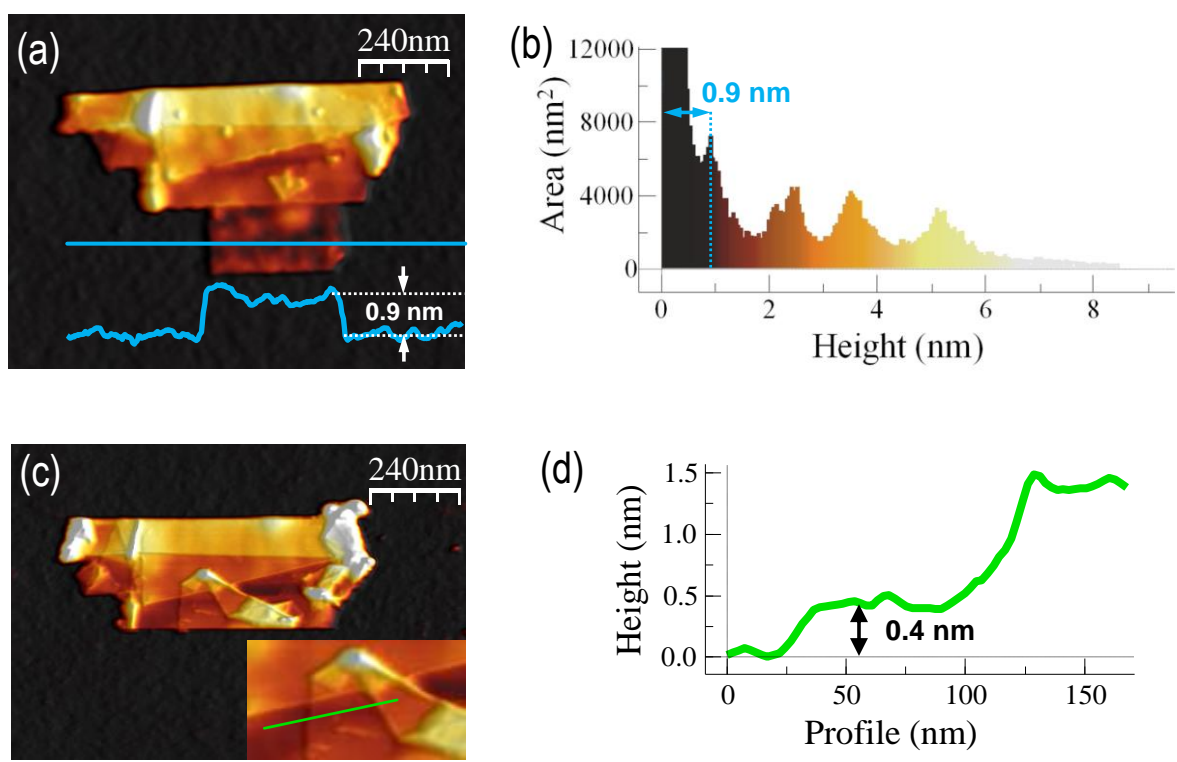

**Figure 3.** AFM topography images of an antimonene flake with a monolayer terrace at the bottom (a) AFM topography showing a ~ 0.2 μm$^2$ antimonene flake with terraces of different heights. The profile is taken along the blue horizontal line in the image. (b) Height histogram of the image in (a) where the different thicknesses of the terraces can be readily seen. For the sake of clarity, the substrate peak has been cut to 12000 nm$^2$. The minimum step height is ~ 0.9 nm compatible with a single layer of antimony adsorbed on the presence of water layers. (c) Same flake as in (a) but after a nanomanipulation process. The lower terrace of the flake was folded upwards resulting in an origami structure with different folds. The inset corresponds to the area of the origami where the lowest step height is found. (d) Profile along the green line in the inset in (c). The lowest step height is ~ 0.4 nm corresponding to a single layer antimonene.

We have also performed basic electrical characterization on few layer antimonene flakes using Conductive-AFM (Supporting Information). Fig. S7d depicts Current *vs*. Voltage (IV) characteristics taken on few layer antimonene flakes (down to 6 nm thickness). The linear dependence of the IV curves is in good agreemenet with theoretical calculations for thicknesses above 2 layers. This result confirms again ambient stability of antimonene.

**Fig. 4** further confirms the effects of water in the flakes. The figure shows three AFM topographic images of the same flakes; Fig. 4a was obtained immediately after sample preparation, Fig. 4b two months later storing the sample under ambient conditions and Fig. 4c was obtained immediately after Fig. 4b but with the sample immersed in water. The inset in Fig. 4c depicts a high-resolution image in one of the flakes acquired while the sample was in liquid. The atomic periodicity is again compatible with that of antimony confirming again the low reactivity of the flakes with water. Fig. 4d shows the profiles along the lines in a-c. No significant differences can be seen between the three topographies.

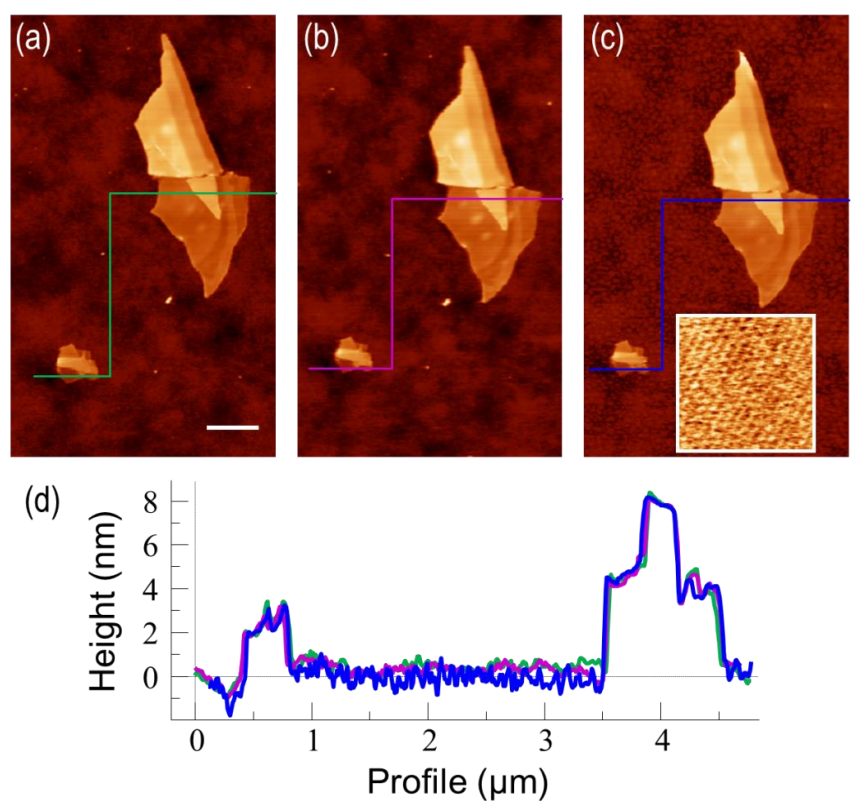

**Figure 4.** AFM topography images of antimonene flakes showing environmental stability of antimonene. (a) Image taken immediately after exfoliation. White scale bar 500 nm. (b) Same as in (a) but two months later. (c) Image taken immediately after (b) but with the sample immersed in water. The inset shows atomic periodicity compatible with antimonene atomic lattice. The selected region is the same as in Fig. 2c. (d) Profiles taken along the lines drawn in a-c. Please notice the similarity of the corrugation on the antimonene flakes confirming again the absence of environmental degradation.

Nevertheless, while imaging in water we observed a tendency of the smallest flakes to exfoliate when the AFM tip was scanned in contact with the flakes. This suggests that water exfoliation assisted by ultrasounds can be a feasible way to obtain thin layer flakes of antimony (work in progress).

To sum up, using mechanical exfoliation combined with a controlled double step transfer procedure we demonstrate that single layers of antimony can be readily produced. These flakes are not significantly contaminated upon exposure to ambient conditions and they do not react with water. DFT calculations confirm our experimental observations and predict a band gap of 1.2-1.3 eV (ambient conditions) for single layer antimonene, which is smaller than that calculated under vacuum conditions at 0 K. Our work confirms antimonene as a highly stable 2D material with promising relevant applications in optoelectronics.

*Experimental*

*Antimonene mechanical exfoliation*: bulk, commercially available antimony material (99.9999%, Smart Elements) was used. Preparation of isolated single-layer and few-layer antimonene flakes was carried out using a modified mechanical exfoliation technique outlined in ref. [9]. By employing a viscoelastic stamp (Gelfilm® from Gelpak®) as an intermediate substrate for exfoliation, thin antimonene flakes were transferred to a Si substrate (with 300 nm of $SiO_2$ capping layer).

*Atomic force microscopy (AFM) imaging*: AFM measurements were carried out using a Cervantes Fullmode AFM from Nanotec Electronica SL. WSxM software (www.wsxmsolutions.com) was employed both for data acquisition and image processing.[14] All the topographical images shown in this work were acquired in contact mode to avoid possible artifacts in the flake thickness measurements.[15] OMCL-RC800PSA cantilevers (probe.olympus-global.com) with a nominal spring constant of 0.39 N/m and tip radius of 15 nm were employed. Low forces of the order of 1 nN were used for imaging to ensure that the flakes would not be deformed by the tip.

*Transmission Electron Microscopy (TEM)*: images were obtained in a JEOL JEM 2100 FX TEM system with an accelerating voltage of 200 kV. The microscope has a multiscan charge-coupled device (CCD) camera ORIUS SC1000 and an OXFORD INCA X-Ray Energy Dispersive Spectroscopy (XEDS) microanalysis system.

*DFT Calculations:* Perdew-Burke-Ernzerhof functional (PBE)[16] was used for the geometry optimization and for the molecular dynamics simulations, and the Heyd-Scuseria-Ernzerhof (HSE06) functional[17] for computing the band gap. Weak interactions were taken into account employing the DFT-D2 method of Grimme.[18] The interaction between ions and electrons has been described by the projector-augmented wave (PAW) method,[19] using plane wave basis sets for the description of the electronic wave function with a cut-off kinetic energy of 700 eV, and imposing periodic boundary conditions. The size of the supercell used is shown in Fig. 2f. In the simulations including solvent effects we performed a previous thermalization at T=298 K using the algorithm of Nosé.[20] All calculations were performed with the Vienna Ab initio Simulation Package (VASP).[21]

*Acknowledgements*
This work was supported by MINECO projects Consolider CSD2010-00024, MAT2013-46753-C2-1 and 2, FIS2013-42002-R and CTQ2013-43698-P, CAM project NANOFRONTMAG-CM (ref. S2013/MIT-2850). The authors acknowledge the allocation of computer time at the Centro de Computación Científica at the Universidad Autónoma de Madrid (CCC-UAM) and the Red Española de Supercomputación (RES). S.D.-T. acknowledges support from the 'Ramón y Cajal' programme. Spanish Ministry of Economy and Competitiveness through The "María de Maeztu" Programme for Units of Excellence in R&D (MDM-2014-0377). The authors acknowledge Dr. C. Munuera for her kind support on gold substrates preparation.

# Supporting Information

**Mechanical Isolation of Highly Stable Antimonene under Ambient Conditions**

By *Pablo Ares*, *Fernando Aguilar-Galindo*, *David Rodríguez-San-Miguel*, *Diego A. Aldave, Sergio Díaz-Tendero*, *Manuel Alcamí*, *Fernando Martín*, *Julio Gómez-Herrero\** and *Félix Zamora\**

## MATERIALS AND METHODS

**Luminescence/Raman spectra** were performed using a WITEC/ALPHA 300RA Raman confocal microscope (Witec GmbH, Ulm, Germany) at ambient conditions. The laser wavelength and power were 532 nm and 1 mW respectively.

## ATOMIC FORCE MICROSCOPY

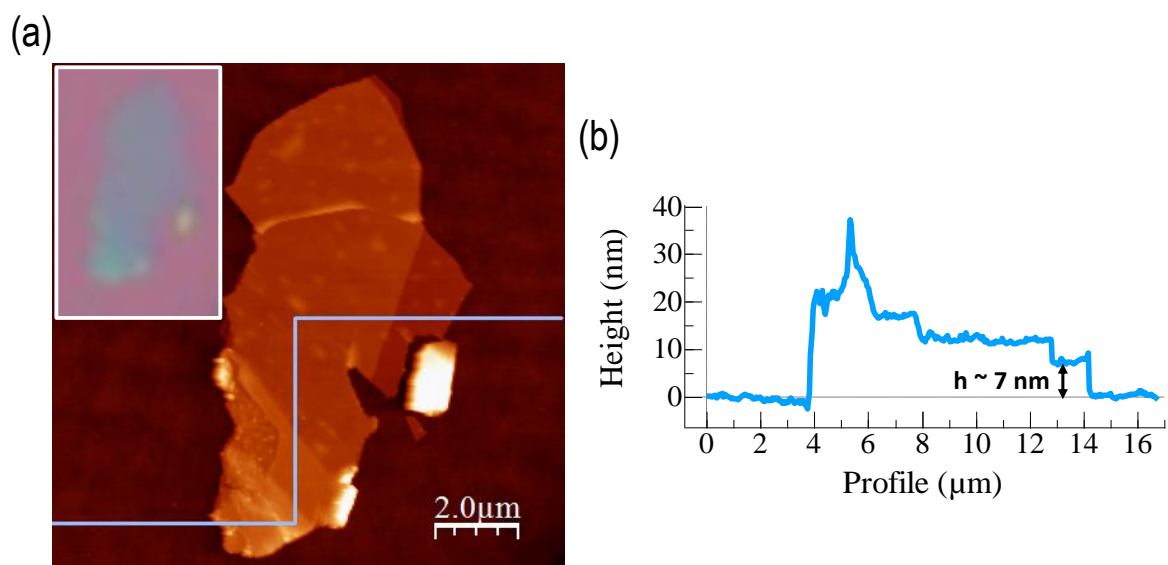

**Figure S1.** (a) AFM topography image of the blueish flake in Fig. 1b in the main text. Inset: Digital magnification from Fig. 1b showing the blueish flake. (b) Height profile along the line in (a) showing the heights of the different terraces in the flake.

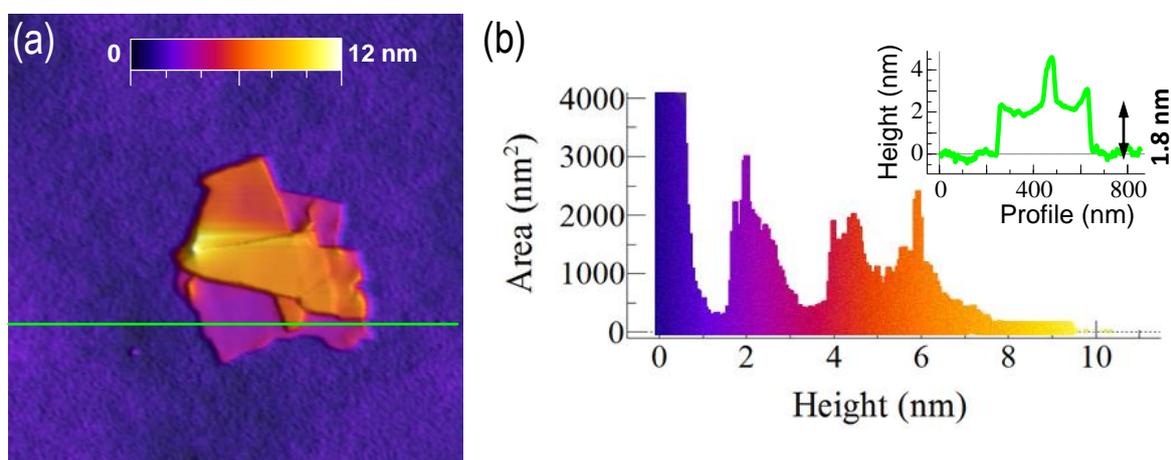

**Figure S2.** Few-layer antimonene flake. (a) AFM topography showing a ~ 0.2 μm$^2$ antimonene flake with terraces of different heights. (b) Height histogram of the image in (a) where the different thicknesses of the terraces can be readily seen. For the sake of clarity, the substrate peak has been cut to 4000 nm$^2$. The inset is a profile along the horizontal line in (a). The minimum step height is ~ 1.8 nm compatible with 2 layers of antimonene.

## PHOTOLUMINESCENCE RAMAN SPECTROSCOPY

Figure S3 depicts a Photoluminescence (PL) Raman spectroscopy study of antimony flakes of different heights on a $SiO_2$/Si substrate. Fig. S3a,b shows optical microscopy images where a variety of flakes of different heights (measured by AFM) and lateral dimensions can be observed. Figure S3c shows PL-Raman spectra taken on different spots marked with crosses in (a,b). We can observe that for bulk antimony (height > 1000 nm) the spectrum matches previous results for Raman bulk antimony[1] with almost no signal coming from the substrate (~ 520 $cm^{-1}$). As the thickness of the flakes decreases (h = 100 nm) the signal from the substrate is becoming dominant and for lower heights the antimony signal is lost. We notice that this behavior is related to the thickness of the flakes, not with their lateral dimensions. Indeed, the h = 100 nm flake is clearly much smaller than the h = 11 and 30 nm ones, but whereas the h = 100 nm spectrum shows the characteristic peaks corresponding to antimony the h = 11 and 30 nm spectra do not.

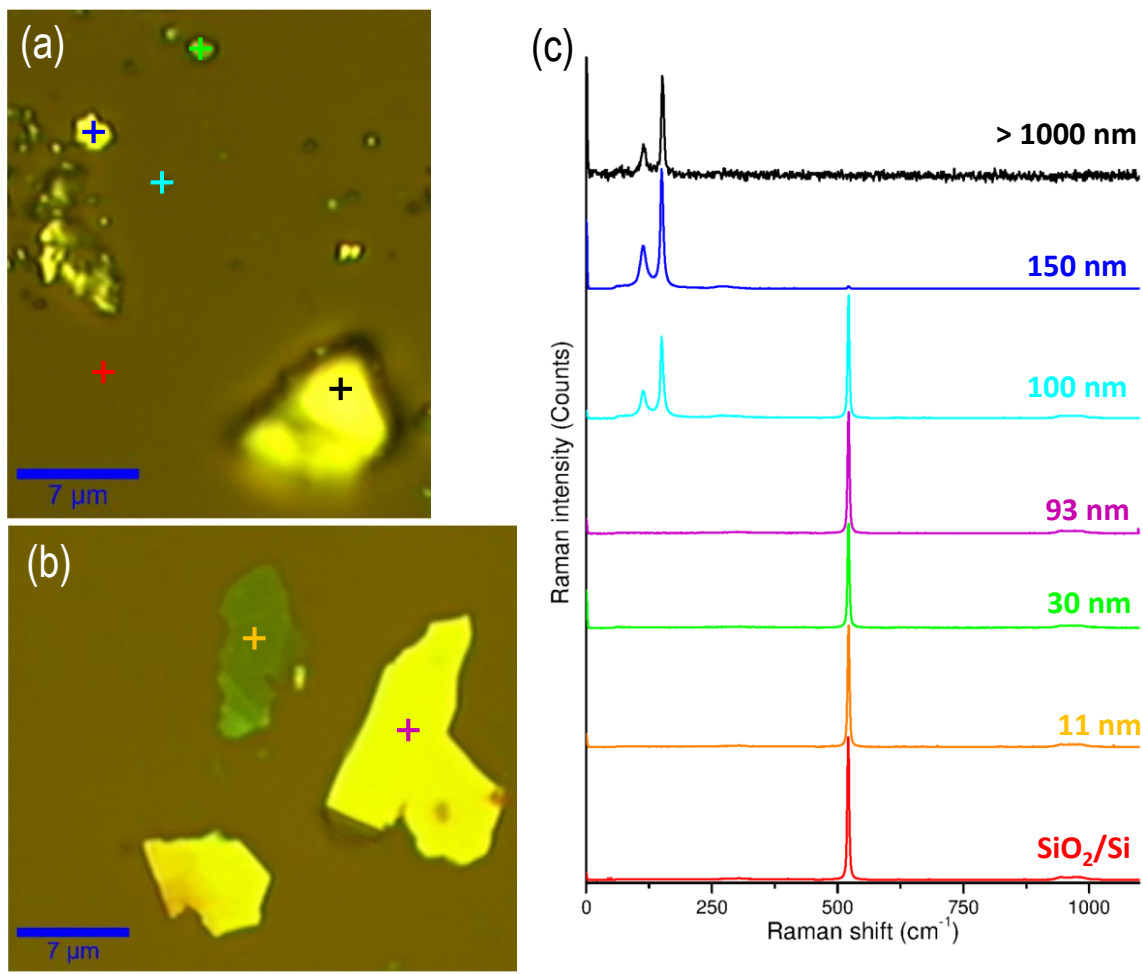

**Figure S3.** (a,b) Optical microscopy images of antimony flakes with a wide range of heights. (c) PL-Raman spectra from the points indicated by crosses in (a,b). The red spectrum corresponds to the substrate; the remaining spectra correspond to antimony flakes of different thicknesses (measured by AFM). Black, h > 1000 nm (bulk antimony); dark blue, h = 150 nm; cyan, h = 100 nm; pink, h = 93 nm; green, h = 30 nm; orange, h = 11 nm; red: $SiO_2$/Si substrate.

# STRUCTURAL CHARACTERIZATION

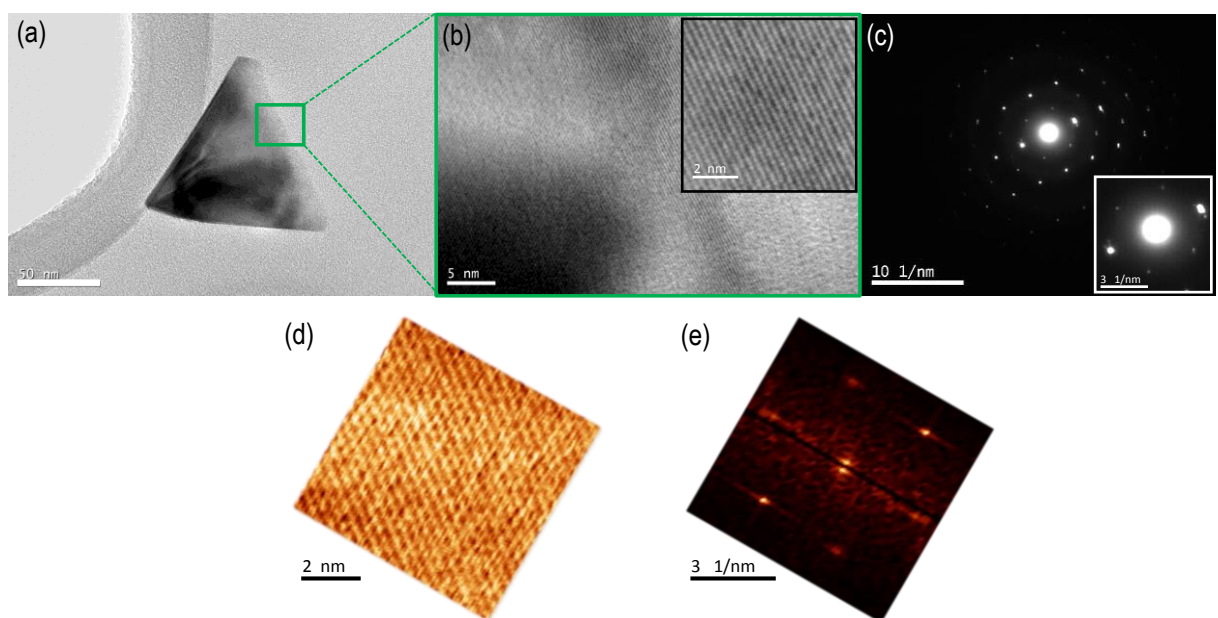

**Figure S4.** (a) TEM image of a thin antimony flake. (b) Magnification of the green area in (a). Inset: Digital magnification of a region in (b). (c) Electron diffraction pattern. Inset: Magnification of the central region. (d) AFM showing atomic periodicity. The image is rotated to match the orientation as in TEM images. (e) Fast Fourier Transform (FFT) image taken from (d) showing the agreement with the electron diffraction pattern.

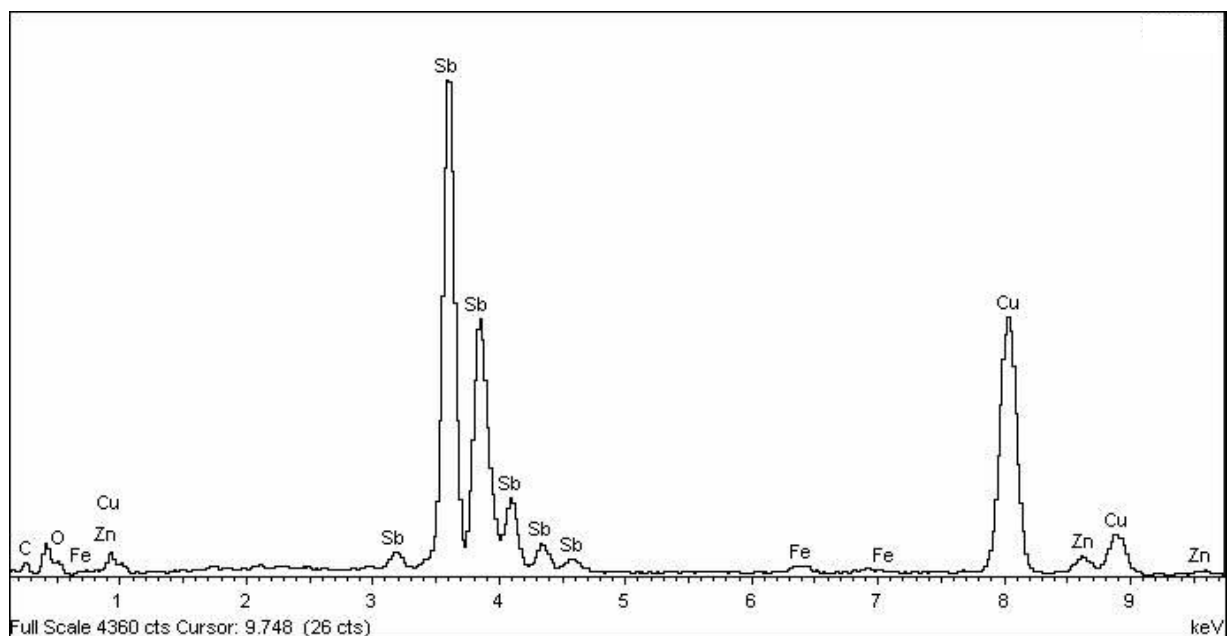

**Figure S5.** X-Ray Energy Dispersive Spectroscopy (XEDS) microanalysis.

**COMPUTATIONAL DETAILS AND CONVERGENCE IN THE SIMULATIONS**

The simulations were carried out in the framework of Density Functional Theory (DFT). In particular we have employed the generalized gradient approximation (GGA) with the Perdew-Burke-Ernzerhof functional (PBE)[2] for geometry optimizations and molecular dynamics simulations, and the Heyd-Scuseria-Ernzerhof (HSE06) functional[3] for computing the band gap. The PBE and HSE functionals have been proved to accurately predict geometries (see e.g. [4]) and electronic properties (see e.g. [5]), respectively. We have taken into account weak interactions, which are missing in most GGA-like functionals, by using the DFT-D2 method of Grimme.[6] The latter includes van der Waals (vdW) contributions through a semiempirical potential added to the total DFT energy. In the DFT calculations, the interaction between ions and electrons has been described by the projector-augmented wave (PAW) method,[7] using plane wave basis sets for the description of the electronic wave function and imposing periodic boundary conditions. The Brillouin zone was sampled with a single Gamma Point. As a check of the method, we have calculated the lattice constants of bulk Sb by using the PBE functional, leading to a=b=4.28Å and c=11.18Å, in reasonable agreement with the experimental values a=b=4.31Å c=11.27Å.[8] For the 1ML and 2ML Sb calculations, we have imposed a cut-off kinetic energy for the plane wave basis of 700 eV. The periodic supercell we have used consists of a slab constructed with one or two atomic monolayers of antimony in a unit cell of (3x3) atoms (**Fig. 2** for the monolayer case). This supercell has hexagonal symmetry with a lattice vector of 12.84 Å, and a vacuum/solvent distance of ~30 Å. Such a large supercell is required for a proper description of the solvent effect, mimicking the experimental conditions. In the geometry optimization with PBE and in the single point energy simulations with HSE06, convergence of the electronic self-consistent procedure was assumed when the energy difference between two consecutive cycles was smaller than $10^{-5}$ eV. Convergence in the geometry optimizations was assumed when all Hellmann-Feynman forces were smaller than 0.02 eV/Å. In the simulations including solvent effects, we explicitly introduce water molecules with the same density of liquid water at room temperature and atmospheric pressure conditions. In these simulations, a previous thermalization was carried out with ab initio molecular dynamics in the canonical ensemble by using the algorithm of Nosé,[9] ensuring a constant temperature. We run the thermalization during ~1 ps at T=298 K with a time step of 0.2 fs. Spin non-polarized molecular dynamics simulations were performed by using an electronic convergence criterion of $10^{-4}$ eV. Since the curves obtained in the computed density of states present a broad profile, band gaps were computed as the energy differences between the Fermi energy ($E_F$) and the lowest energy in the conduction band. Figure S6 shows the calculated density of states from which the value of the gap has been extracted. All calculations were performed with the Vienna Ab initio Simulation Package (VASP).[10]

We have carefully checked convergence of the results with respect to the different parameters used in the calculations. We summarize below the different checks that have been performed.

<u>-Kinetic energy cutoff in the plane wave expansion:</u>
Although valence electrons in Sb atoms can be reasonably described by using a cut off of ~300 eV, oxygen atoms require a higher value. A cut off of 500 eV is usually enough for a correct description of individual or a few water molecules. However, since the dielectric constant of liquid water is a property that is rarely studied by explicitly including individual molecules in fully quantum-mechanically calculations, it is not possible to assure a priori that this value provides an accurate enough description. This is why we have carried out calculations not only with 500 eV but also with 700 eV for cut-off energy. The results for the band gap in the one-monolayer case including solvent effects show that the kinetic energy cutoff is practically converged by using a cut off of 500 eV (at least for the number of water molecules included in our simulations, see below):

                  Cutoff: 500 eV – Band gap: 1.147 eV
                  Cutoff: 700 eV – Band gap: 1.148 eV

- Number of water molecules:

To determine how many water molecules are required to reproduce the dielectric constant of liquid water, we have performed calculations by using different thicknesses, namely ~30 Å (125 water molecules) and ~24 Å (100 water molecules). We have found that a thickness of 24 Å is good enough to converge the bulk dielectric constant of the liquid water:

                  100 molecules – Band gap: 1.147 eV
                  125 molecules – Band gap: 1.150 eV

- Number of bands:

In DFT simulations, increasing the number of bands in the simulations can also affect the value of the band gap. This is why it is necessary to check converge with respect to the number of bands. To this end, we have carried out two different calculations: one including 896 bands (default number of bands in VASP for the number of electrons in the system within the spin polarized formalism we have employed) and 960 (i.e. increasing by ~10%). The conclusion of this test is that 896 bands is enough to describe the density of states near the Fermi level and to obtain a proper description of the band gap:

                  896 bands – Band gap: 1.15001 eV
                  960 bands – Band gap: 1.14999 eV

- K-Points:

We have sampled the Brillouin zone by using different numbers of K-points in the three directions ($M$ x $M$ x $M$ K-points). Since the computational effort that would be required to perform a fine K sampling in the system that explicitly includes the solvent is unaffordable, we have performed such convergence tests for the case of a single monolayer of Sb in vacuum. To this we have used a Monkhorst-Pack grid.[11] The results are:

                  K-points: 1x1x1 – Band gap: 1.595 eV
                  K-points: 3x3x1 – Band gap: 1.543 eV
                  K-points: 5x5x1 – Band gap: 1.541 eV
                  K-points: 7x7x1 – Band gap: 1.541 eV

Since the different between the 7x7x1 sampling and the 1x1x1 (only Γ point) is just a few meV, we can consider that the Γ-point sampling is reasonable to study the effect of the water solvent on the Sb.

Finally, the evaluation of the band gap was carried out by assuming different situations, increasing the degree of complexity, and evaluating different factors that might modify it under ambient conditions, for the 1ML case:

(i) We first assumed the crystal structure of Sb and we optimized the geometry of the monolayer in gas phase at 0 K.
(ii) In order to understand the effect of temperature we run a second set of simulations on the monolayer employing ab initio molecular dynamics assuming T= 298 K during 1 ps.
(iii) The effect of the solvent was taken into account running ab initio molecular dynamics with T= 298 K during 1ps and including explicitly the water molecules.

(iv) We also evaluated the effect of the presence of oxygen including explicitly $O_2$ molecules and performing ab initio molecular dynamics at T= 298 K during 1ps.

We then computed the band gap in single point simulations at the HSE level, by using an averaged geometry over the last 50fs of each dynamics. Table S1 summarizes the results obtained for the band gap within the four above-mentioned scenarios.

**Table S1.** Band gap obtained within the HSE method using different geometries and situations.

| 1ML | Band gap (eV) |
|---|---|
| Vacuum – 0K | 1.59 |
| Vacuum – 298K | 1.28 |
| Water – 298K | 1.22 |
| Oxygen – 298K | 1.29 |

Notice that the computed band gap in vacuum at 0K is in very good agreement with that obtained by Akturk *et al* [12] and is 0.7 eV smaller than that given in ref. [13]. A possible reason for the difference might be the slightly different geometries employed in the three works.

As can be seen, the effect of increasing temperature from 0 to 298K is to reduce the bandgap by approximately 0.3 eV and that of adding liquid water at room temperature is to reduce it a bit more (approximately 0.07 eV). Addition of oxygen does not have any observable effect. Consequently, the most likely value of the bandgap at ambient conditions lies within the interval 1.2-1.3 eV.

For the two monolayers case we only consider the two limit cases: vacuum at 0K and water at 298K.

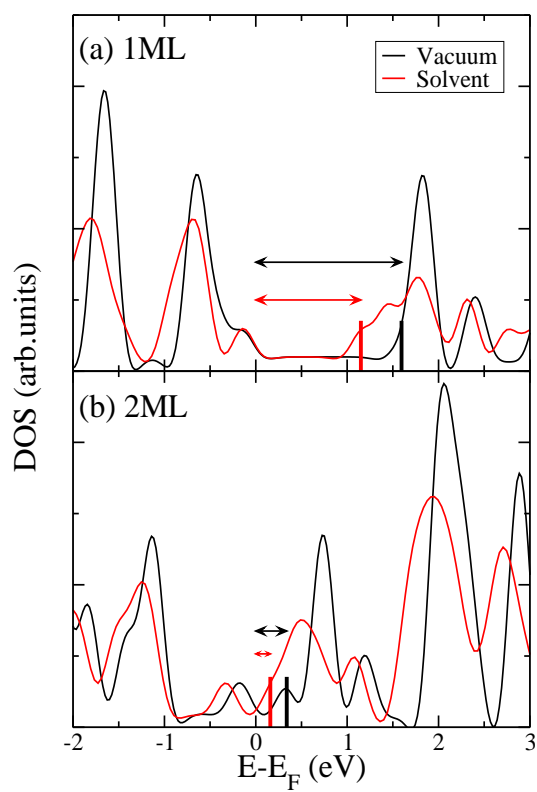

**Figure S6**. Calculated densities of states for 1ML and 2ML antimonene in vacuum at T=0K (black curves) and in the presence of a water solvent at T=298K (red curves) as a function of electron energy (referred to the Fermi energy, $E_F$).

# ELECTRICAL CHARACTERIZATION OF FEW LAYER ANTIMONENE FLAKES

As an additional proof of the nature of our antimony flakes we have transferred few layer antimonene flakes to gold on borosilicate glass substrate. As in the case of graphene and other 2D materials, optical microscopy fails detecting very thin layers on metals, hence the minimum thickness we could detect was about 6 nm. Using a metallized AFM tip we have electrically contacted the thinnest part of the flake. As expected, the conductance on the gold substrate is higher than in the few layer flake, but this one is still very high. We have performed this experiment after exposing the sample to ambient conditions during several days confirming the ambient stability of antimonene. The electrical characterization was also performed in ambient conditions. Both IV curves, on gold and on few layer antimonene, show a clear linear dependence. This is obviously expected for gold but also for our flake (which has more than 2 layers) because for this thickness calculations predict a metallic behavior.

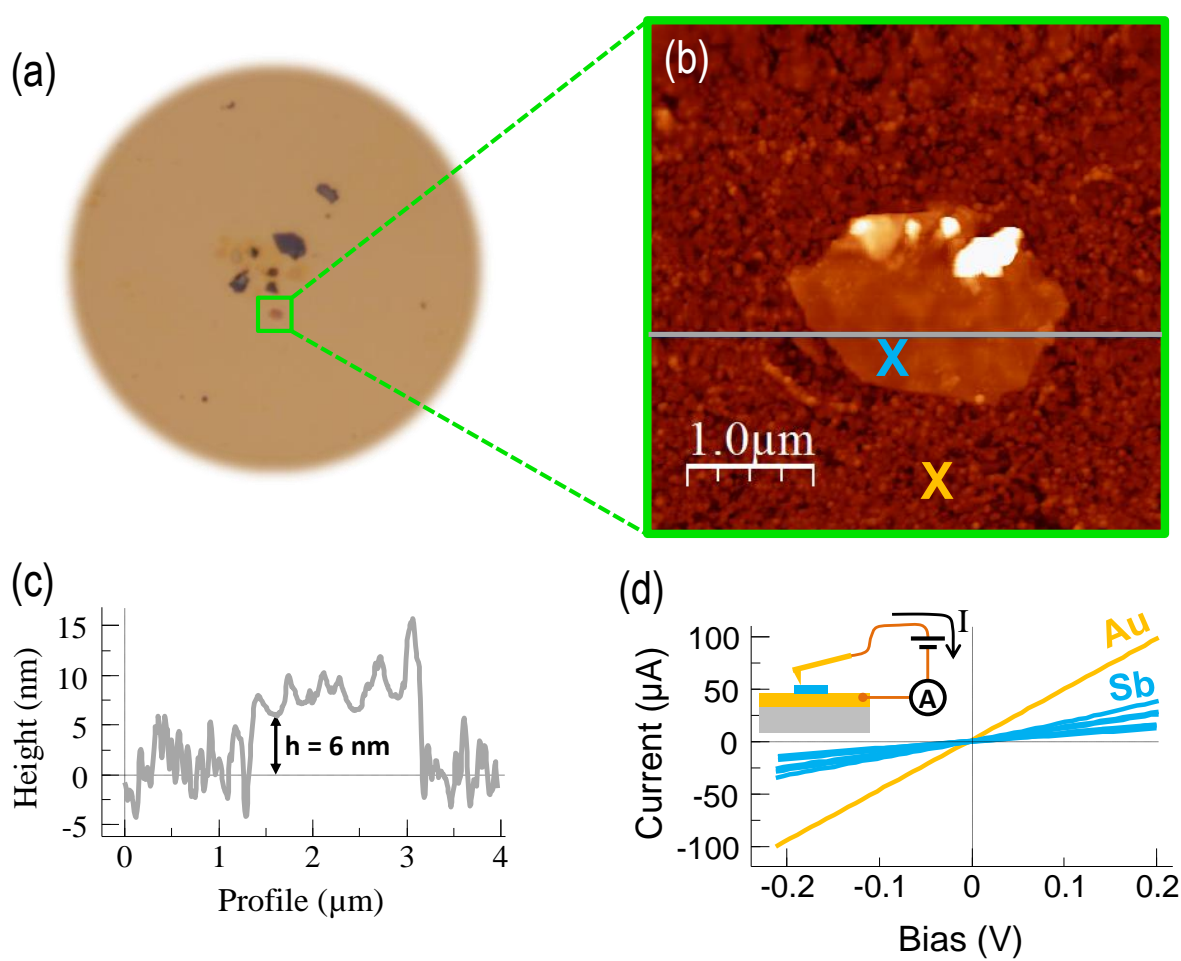

**Figure S7**. Electrical characterization of few layer antimonene. (a) Optical microscopy image showing a thin antimonene flake enclosed by a green square. (b) AFM topography image showing the aforementioned flake. (c) Profile along the horizontal gray line in (b). The minimum height of the flake is 6 nm. (d) Current *vs.* Voltage characteristics taken on gold (Au, orange cross in (b)) and in several few layer antimonene flakes with thickness down to 6 nm (Sb, blue cross in (b)). The slight differences in slopes in the few layers antimonene flakes IVs are related with different tip-sample contact areas. The inset is a schematic representation of the electrical circuit.